# Konstantin Kostenko

Innopolis University

# Real-time Communication between Robot PLC and PC over Ethernet-based Protocols

*The article provides a comparative analysis of four communication protocols between Omron PLC and PC for their use in real-time control of an industrial robot. The need for real-time communication appears when the robot control system is located on an external PC. The subject of the study is the FINS, CIP, UDP, and OPC protocols. The physical medium for data transmission is Fast Ethernet 100 Mbit/s. The average time characteristics of reading, writing, and write/read cycle for each protocol are given. The evaluation of the applicability of protocols for real-time robot control and the convenience of their use is also given.*

## 1. Introduction

The development of robotics leads to the emergence of increasingly sophisticated control algorithms that allow the robot to perform complex and highly accurate operations. Robots are usually equipped with programmable logic controllers (PLCs), where the hardware and software are specially adapted to the industrial environment. PLC is a cost-effective and reliable solution for controlling complex systems. However, modern sophisticated algorithms require significant computing power, which robot PLCs usually do not have. The solution in this situation is the use of an external control system running on a PC. In this case, it is necessary to provide reliable and high-speed communication between both control systems.

This article will discuss the use of several real-time communication protocols supported by the Omron NJ501-1300 controller. This PLC is used in the control system of a cable robot at the Robotics Development Center of Innopolis University [1, 2]. The robot has a reconfigurable design with a number of cables from 4 to 12 with Omron servo motors in the winches.

## 2. Real-Time Communication between Robot and PC

In studies related to robot control, it is often necessary to implement various schemes and control algorithms. A convenient tool for creating such algorithms is Matlab, which has earned recognition among researchers. Unlike programming languages of industrial controllers, Matlab, working on the computer, provides a wide range of built-in functions and algorithms that allow one to implement sophisticated algorithms in the shortest possible time.

The robot control system is divided into two levels: the lower and upper. The lower level implemented in the robot PLC provides basic control functions such as servo control, state control, etc. The upper level implemented on the PC is responsible for the formation of the control flow, which includes the functions of global planning, feedback analysis, state correction, etc. Thus, there is a need for real-time communication between the robot controller and the application in Matlab running on the computer.

Many articles are devoted to the communication between PLCs and PCs. For communication between PLCs, specially developed fieldbus protocols are used (EtherCat, Profinet, Modbus, EtherNet/IP, etc.), which provide real-time communication with delays of the order of several microseconds [3, 4, 5, 6, 7]. In the case of connection of a PC to these fieldbuses, delays associated with the operation of the computer OS in multitasking mode emerge, which does not guarantee the execution of the program code in a certain time. The integration of PCs and PLCs is dealt with in [8, 9, 10], which, however, do not provide a reference to the values of time delays that arise during communication. Communication between the application in Matlab and the PLC is also considered in [11, 12]. Due to the lack of comparative information on the time delays occurring in the real-time communication between Omron PLC and PC over various protocols, this study was conducted.

As the research objective, four communication protocols of the robot PLC with an external PC are selected: FINS, CIP, UDP, and OPC. The first three are supported directly by the Omron controller, and the latter works as an additional layer over the CIP protocol in the computer OS. The consideration of OPC is due to its popularity in the environment of industrial automation. The physical medium for data transmission is Fast Ethernet 100 Mbit/s.

### 2.1. Hardware and software components

The Omron NJ501-1300 PLC with firmware version 1.12 is used in the robot. The robot is controlled from the HP Pavilion 15-p158nr laptop PC with a dual-core Intel Core i7-4510U 2.0 GHz and 12 Gb RAM.

The software running on the PC is shown in Table. 1.

Table. 1. Software of the test PC.

| | |
|---|---|
| Microsoft Windows 10 Pro Edu | |
| Omron CX-Compolet 1.7 | The library of .NET assemblies that implements the communication between the PC application and the robot PLC using the FINS and CIP protocols. |
| Omron Sysmac Gateway 1.7 with FINS Gateway | Sysmac Gateway is a communications middleware that provides a factory automation network environment on a computer. |
| Kepware KEPServerEX 6 | OPC Server software |
| Omron Network Configurator for EtherNetIP 3.61 | A network utility that allows one to create Tag Data Links between the PLC and the PC on the network. |
| Matlab R2015b 32-bit | The Matlab version is determined by using the 32-bit CX-Compolet library. The most recent 32-bit version of Matlab is selected. |

To test the communication time parameters in the robot PLC program, two global variables of the type LREAL (double) are created: CIn and COut. The variable CIn is intended for receiving data (has Input Network Publish attribute), and COut - for sending (has Output Network Publish attribute). Every program cycle (Task Period = 1 ms) the value of CIn variable is assigned to COut variable, so that the received data is immediately put in the send buffer. Similarly, one can transfer data sets of several tens or hundreds of elements, but this is not necessary for this test.

### 2.2. FINS

Factory Interface Network Service (FINS) [13] is a network protocol used by Omron PLCs. The FINS communications service has been developed by Omron to provide a consistent way for the PLCs and computers on the Omron factory automation (FA) networks to communicate in order to send and receive data, change modes, and so on. It provides the following features [14]:

• Communications instructions are executed in the user program.

• Writing data, changing modes, reading detailed information about Units, and so on, can be executed without any particular knowledge of communications.

• Units and Boards that support FINS commands return responses automatically, so there is no need for a program at the receiving end.

• The FINS communications service is mainly used between Omron CPU Bus Units, CPU Units, and Support Boards for FA Computers. By correctly setting up information such as headers, however, it can also be used from ordinary Ethernet communications devices.

The PC communicates with the PLC over the FINS protocol by reading/writing the PLC memory areas with which the global variables of its software are associated. CX-Compolet and Sysmac Gateway must be installed on the PC so that it can communicate over FINS. This software is recommended and distributed by Omron as a standard solution. The application in Matlab communicates with the PLC using .NET assembly FgwCompolet, part of CX-Compolet. To send a request to read the memory area of the PLC, it is necessary to generate a text command with the code '0101', and to write to the PLC memory area – a command with the code '0102'. The implementation of sending/receiving data in the CX-Compolet assembly is such that one have to convert numeric data into a text string and back, which is not convenient for programming.

When testing communication time characteristics, 100 000 trials of reading, writing and write/read cycle were conducted. The average time of operations is given in Table. 2.

Table. 2. Average read/write time using the FINS protocol.

| Read, ms | Write, ms | Write/Read Cycle, ms |
|---|---|---|
| 4.41 | 4.41 | 5.59 |

After the write/read command is submitted, the PLC returns a response. For the write command, the response is the success/failure status. Taking into account that the connection between the controller and the computer is organized in the laboratory environment and does not experience interference effects, the cable connection is of short length, the probability of non-delivery of packets tends to zero. Therefore, one can not analyze the response with the status of the write request and perform cyclic writing/reading in asynchronous mode. Commands for writing and reading data to/from the PLC were submitted sequentially without delay, and then the responses were read, the first of which related to the write operation was ignored. This approach reduced the write/read cycle time from 4.41 + 4.41 = 8.82 ms to 5.59 ms.

### 2.3. CIP

Common Industrial Protocol (CIP) [15] encompasses a comprehensive suite of messages and services for a variety of manufacturing and process automation applications, including control, safety, synchronization, motion, configuration and information. CIP is a truly media-independent object-oriented protocol that is supported by hundreds of vendors around the world. One of the possible networks is EtherNet/IP, which adapts CIP to Ethernet technology.

Each CIP object has well-defined attributes (data), services (commands), and behaviors (responses to events). CIP's producer-consumer communication model provides more efficient use of network resources than a pure source-destination model by allowing the exchange of information between the sending device (e.g., the producer) and many receiving devices (e.g., the consumers) without requiring data to be sent multiple times by a single source to each individual destination. In producer-consumer networks, a message is identified by its connection ID, not its destination address. The producer-consumer model provides a clear advantage for the users of CIP Networks by making efficient use of network resources in the following ways:

• If a node wants to receive data, it only needs to ask for it.

• If a second (third, fourth, etc.) node wants the same data, when it asks for it, the device will provide the multicast address and Connection message from the network.

The connection between a producer and consumer (called Tag Data Link) is made using the Network Configurator and Sysmac Gateway. The application in Matlab communicates with the PLC using the .NET VariableCompolet and NJCompolet assemblies, parts of CX-Compolet.

The link mechanism is based on variables (Tags) that have an input or output attribute. Two variables: CIn and COut are created in Sysmac Gateway Console application, the type and purpose of which are the same as the previously described variables CIn and COut in the robot's PLC. The variable CIn of the controller is linked with the COut variable of the PC, and the CIn variable of the PC is linked with the controller's COut variable. The Request Packet Interval (RPI) is set for both links to 1 ms. This interval provides the maximum update rate of the links. The application in Matlab can write/read the values of the CIn and COut variables of the Sysmac Gateway server, which, in its turn, maintains the relevance of their values using the created links.

When testing communication time characteristics, 100 000 trials of reading, writing and write/read cycle were conducted. The average time of operations is given in Table. 3.

Table. 3. Average read/write time using the CIP protocol.

| Read, ms | Write, ms | Write/Read Cycle, ms |
|---|---|---|
| 4.07 | 3.92 | 4.00 |

The read and write operations were performed using the NJCompolet assembly of the CX-Compolet library. To test cyclic writing/reading, Tag Data Links have been created between the PC's and the robot PLC's variables. The Sysmac Gateway variables were written and read using the CX-Compolet's VariableCompolet assembly. The link mechanism allows one to write/read data to/from the robot controller in asynchronous mode, which reduces the write/read cycle time by half.

### 2.4. UDP

The UDP (User datagram protocol) protocol [16] is one of two popular protocols used on the transport layer of the OSI model, along with the TCP protocol. There are common transport-layer tasks that are handled by both TCP and UDP – segmentation of data coming from the application layer and addressing applications (sending and receiving) via ports. In addition, TCP provides functions, such as reliable delivery and connection setup, which UDP does not provide. The main purpose of UDP is the fastest delivery, so UDP is the thinnest possible interlayer between the network layer and the application layer.

The Omron NJ501 controller supports both UDP and TCP, but due to conducting the experiment in ideal laboratory conditions, in which there is no need to control the delivery of messages, UDP was selected for the tests.

Since UDP is the basic transport protocol, additional effort is needed to program communication over UDP in the PLC. SktUDPCreate, SktUDPRcv, and SktUDPSend function blocks (FB) have been used for this purpose. When testing cyclic writing/reading, there were two receive/send rungs where SktUDPRcv and SktUDPSend were connected sequentially so that after receiving a message, it was immediately sent back. Two rungs were used due to the fact that the execution time of each FB can be of several Task Periods, while receiving and sending should be performed as soon as possible. Thus, these two rungs were executed alternately, receiving and sending back the received messages.

The application in Matlab supports UDP communication through the UDP object of the Instrument Control Toolbox.

When testing communication time characteristics, 100 000 trials of reading, writing and write/read cycle were conducted. The average time of operations is given in Table. 4.

Table. 4. Average read/write time using the UDP protocol.

| Read, ms | Write, ms | Write/Read Cycle, ms |
|---|---|---|
| 1.78 | 2.00 | 4.00 |

The implementation of communication using the UDP protocol required the greatest amount of programming effort among all the protocols examined. Due to the peculiarities of UDP implementation in the Omron controller, we could not achieve read/write asynchrony. Therefore, the cyclic writing/reading was performed synchronously.

### 2.5. OPC

OPC (Open Platform Communications) [17] is the interoperability standard for the secure and reliable exchange of data in the industrial automation space and in other industries. It is a platform for independent and guaranteed the flow of information among devices from multiple vendors. The OPC Foundation is responsible for the development and maintenance of this standard.

The OPC standard is a series of specifications developed by industry vendors, end-users and software developers. These specifications define the interface between Clients and Servers, as well as Servers and Servers, including access to real-time data, monitoring of alarms and events, access to historical data and other applications.

To set up communication between the application in Matlab and the robot PLC using the OPC protocol, it is necessary to use server software. Kepware KEPServerEX is recommended by Omron for connecting PLCs with the computer via the OPC standard. KEPServerEX contains a set of drivers for communication with various PLCs. The driver for the Omron NJ501 controller is based on the CIP protocol.

The server needs to be configured to establish a connection to the PLC. Data flow has the following structural hierarchy: Channel – Device – Item. When creating a Channel, the network interface of the PC and the driver for a corresponding PLC are selected. When creating a Device, the communication parameters with the PLC are specified and there is an option of automatic Items creation for all global variables in the PLC for which the Network Publish attribute is set. Item is a container of transmitted information via the OPC protocol and is similar to CIP Tag. For the tests, the parameter Scan Mode was set to Request all data at scan rate = 10 ms, which is the minimum data polling interval.

The application in Matlab supports OPC communication through the Data Access Client Object of the OPC Toolbox.

When testing communication time characteristics, 100 000 trials of reading, writing and write/read cycle were conducted. The average time of operations is given in Table. 5.

Table. 5. Average read/write time using the OPC protocol.

| Read, ms | Write, ms | Write/Read Cycle, ms |
|---|---|---|
| 15.63 | 7.82 | 15.64 |

Matlab OPC Toolbox and KEPServerEX allow to read or write items both synchronously and asynchronously. In the case of asynchronous read or write operations, a callback is generated when the operation is completed. To reduce the cyclic writing/reading time, an asynchronous write operation was used, the callback of which was not processed. Then a synchronous read operation was performed. Thus, the write/read cycle time was kept at the single reading time.

### 3. Discussion

When examining the communication capabilities of the HP laptop and the Omron NJ501 controller, the following comparative results for the read/write times over various protocols were obtained (Table. 6):

Table. 6. Average read/write time for various protocols.

| Protocol | Read, ms | Write, ms | Write/Read Cycle, ms |
|---|---|---|---|
| FINS | 4.41 | 4.41 | 5.59 |
| CIP | 4.07 | 3.92 | 4.00 |
| UDP | 1.78 | 2.00 | 4.00 |
| OPC | 15.63 | 7.82 | 15.64 |

Due to the fact that communication via the OPC protocol passes through an additional communication layer, the time characteristics of the data exchange are the worst of all the protocols considered. Such values do not allow to use it in the task of external robot control in real time. However, it can be used to monitor the state of the robot and turn on/off its functions. Communication via the OPC protocol is conveniently programmed in Matlab.

UDP, in contrast, is the minimum interlayer between the network layer and the application layer of the OSI model. Therefore, it showed minimal delays in reading and writing data between the PLC and the PC. Although it is supported by the Omron controller, additional effort is required to program communication from the controller side. A sophisticated algorithm has been implemented, nevertheless, it was not possible to achieve write and read asynchrony during cyclic transmission due to implementation peculiarities. However, the average cyclic writing/reading time is minimal among all the protocols considered. Communication via the UDP protocol is conveniently programmed in Matlab. An additional advantage of this protocol is the absence of any additional software necessary for communication. The obtained time characteristics allow using the UDP protocol in the task of external robot control in real time. However, its use for monitoring and switching the state of the robot requires the development of additional software interfaces.

FINS is a native Omron protocol, which allows reading/writing the memory addresses of the PLC associated with global variables. No effort is needed to set up communication from the PLC side. However, the implementation of sending/receiving data in the CX-Compolet assembly is such that one have to convert numeric data into a text string and back, which is not convenient in terms of coding. The obtained time characteristics allow using the FINS protocol in the task of external robot control in real time. However, its use for monitoring and switching the state of the robot requires the development of additional software interfaces.

The most convenient for use and demonstrating minimal time delays in cyclic writing/reading was the CIP protocol. The Tag Data Links between variables in the PLC and the PC allow one to perform write and read operations in asynchronous mode. An additional convenience is to access the global variables of the PLC from the application in Matlab by name. The obtained time characteristics allow using the CIP protocol both in the task of external robot control in real time and for monitoring the robot's state and switching on/off its functions.

It should be noted that the obtained communication time characteristics correspond to the test hardware/software configuration and can change in other configurations. Therefore, the obtained values should be perceived as relative when deciding on the choice of protocol in each particular case.

## 4. Conclusion

Four protocols of Omron PLC and PC communication over the Ethernet network in real time were considered: FINS, CIP, UDP, and OPC. The purpose of the study was to determine the possibility of using them to monitor and control the cable robot in real time in the case when the control system is located on a PC, allowing the implementation of sophisticated algorithms. Mainly, the time of the write/read cycle was examined, which directly affects the frequency of control actions. Nevertheless, the convenience of the software implementation of communication over a protocol in the PLC–PC–Matlab bundle was also taken into account.

In our particular case, the choice was stopped on the CIP protocol, which, on the one hand, provides the minimum time delays in cyclic writing/reading, and on the other hand, provides convenient coding in the Matlab environment.

The test results showed that the FINS and UDP protocols can also be used to control the robot in real time. The OPC protocol, however, can only be recommended for monitoring the status of the robot and enabling/disabling its functions.